\documentclass[epj]{svjour}
\usepackage{graphicx}
\usepackage{mathptmx}      % use Times fonts if available on your TeX system
\usepackage{flushend}
\usepackage[numbers,sort&compress]{natbib}
\usepackage[colorlinks,citecolor=blue,urlcolor=blue,linkcolor=blue]{hyperref}
\usepackage{amsmath}
\usepackage{amssymb}
\usepackage{array}
\usepackage{tabularx}
\usepackage{snapshot}

\begin{document}
\title{\boldmath Rare top quark decays at a 100 TeV proton-proton collider: $t \rightarrow bWZ$ and $t\rightarrow hc$.}
\subtitle{}
\author{Andreas Papaefstathiou\inst{1, 2,}\thanks{\emph{e-mail:} apapaefs@nikhef.nl}%
 \and Gilberto Tetlalmatzi-Xolocotzi\inst{2,}\thanks{\emph{e-mail:} gtx@nikhef.nl}%
}                     % Do not remove
\offprints{}          % Insert a name or remove this line
\institute{Institute for Theoretical Physics Amsterdam and Delta
  Institute for Theoretical Physics, University of Amsterdam, Science
  Park 904, 1098 XH Amsterdam, The Netherlands.
          \and
          Nikhef, Theory Group, Science Park 105, 1098 XG, Amsterdam,
          The Netherlands. 
}
\date{Revised version: 5th April 2018}
\abstract{
We investigate extremely rare top quark decays at a future proton-proton collider with centre-of-mass energy of 100 TeV. 
We focus on two decay modes: radiative decay with a $Z$ boson, $t \rightarrow b WZ$, and flavour-changing neutral decay with a Higgs boson,
$t \rightarrow h c$, the former being kinematically suppressed with a branching ratio of 
$\mathcal{O}(10^{-6})$~\cite{Altarelli:2000nt}, and the latter highly loop-suppressed, with a branching ratio of 
$\mathcal{O}(10^{-15})$~\cite{AguilarSaavedra:2004wm}. We find that $t \rightarrow b WZ$ will be very challenging to observe
in top quark pair production, even within well-motivated beyond-the-Standard Model scenarios. For the mode $t\rightarrow h c$ we find a stronger sensitivity than that obtained by any future LHC measurement by at least one order of magnitude. 
%
%\PACS{
%      {PACS-key}{discribing text of that key}   \and
%      {PACS-key}{discribing text of that key}
%     } % end of PACS codes
} %end of abstract

\authorrunning{A. Papaefstathiou and G. Tetlalmatzi-Xolocotzi}
\titlerunning{ Rare top quark decays at a 100 TeV}

\maketitle

\section{Introduction}
The top quark's large mass suggests that it may be intimately-connected to the mechanism of electroweak symmetry breaking (EWSB). 
Furthermore, it is unique as a colour-charged particle because of the fact that it decays before hadronizing. The top
quark decays mostly through the channel $t\rightarrow b W$ with much smaller contributions from other SM processes. 
The QCD production of top quarks at the Large Hadron Collider (LHC) is expected to be high, with
 $\sigma(pp \rightarrow t\bar{t}) \simeq 950$~pb at 14 TeV centre-of-mass energy~\cite{Czakon:2013goa}, and would be 40 times larger at a proton collider
with a centre-of-mass energy of 100 TeV (e.g. the Future Circular hadron-hadron Collider -- FCC-hh~\cite{Mangano:2016jyj, Contino:2016spe}).
Given its large production rates, the top quark provides simultaneously interesting tests of QCD (through its production mechanisms)
and of electroweak physics (through its decay channels). The Tevatron has already measured several properties 
of the top following its discovery, and the LHC is already adding further precision measurements. The purpose of the present study is to 
investigate to which extent a FCC-hh can provide valuable information to our knowledge of the top quark properties and couplings. Here, we focus on the potential for the observability of the radiative SM decay mode, $t \rightarrow bWZ$, and on the decay of the top quark through the direct interaction between the top and charm together with the Higgs boson, i.e. $t\rightarrow h c$.

The decay of the top quark to a bottom quark, a $Z$ and a $W$ boson, $t \rightarrow bWZ$, received some attention 20 years ago, with several studies addressing its rate and potential sensitivity to new physics~\cite{Mahlon:1994us, Jenkins:1996zd, Mahlon:1998fr, DiazCruz:1999mq, Altarelli:2000nt}. A peculiar feature of this process is that it occurs near the kinematical threshold: $m_t \simeq m_b + m_W + m_Z$ and is thus suppressed within the Standard Model (SM), predicted to be $\mathcal{O}(10^{-6})$. In the studies of Refs.~\cite{Mahlon:1994us, Mahlon:1998fr, Altarelli:2000nt}, it was pointed out that one has to take into account the finite widths of the $Z$ and $W$ bosons when calculating this decay mode. Indeed, if the widths are ignored, the current particle data group nominal values~\cite{Olive:2016xmw} 
of the particles involved: $m_t \simeq 173.1$~GeV, $m_b \simeq 4.18$~GeV, $m_Z \simeq 91.19$~GeV and $m_W \simeq 80.39$~GeV 
would imply $m_t < m_b + m_W + m_Z$ and would suggest a kinematically-forbidden decay. Nevertheless, if the gauge boson widths are 
properly taken into account the decay can proceed. This leads to $\mathcal{O}(10^4)$ top quark pair production events containing the decay during 
the lifetime of the high-luminosity LHC (HL-LHC) with an integrated luminosity of $3000$~fb$^{-1}$, inclusively over the 
decays of the $Z$ and $W$ bosons. Consequently, the process will likely be impossible to observe at the LHC. 
However, given the substantial increase in cross section at the FCC-hh this channel is expected to yield $\mathcal{O}(10^6)$ events, justifying a more detailed investigation into its observability.

Flavour-changing neutral (FCN) decays of the top quark appear at one loop and have a strong suppression due to the Glashow-Iliopoulos-Maiani 
(GIM) mechanism and second-third generation mixing~\cite{Eilam:1990zc,, Mele:1998ag, AguilarSaavedra:2004wm, Craig:2012vj}. Within the SM, 
this suppression leads to the following branching ratios: $\mathrm{BR}(t \rightarrow \gamma c) \sim 10^{-14}$, 
$\mathrm{BR}(t \rightarrow   gc ) \sim 10^{-12}$, $\mathrm{BR}(t \rightarrow   Z c) \sim 10^{-14}$ and 
$\mathrm{BR}(t \rightarrow  h c ) \sim 10^{-15}$~\cite{AguilarSaavedra:2004wm}, rendering them unobservable at the current and 
future colliders. Consequently, the measurement of such processes within current capabilities would clearly signal the presence of BSM phenomena. 
Here, we focus on the transition $t \rightarrow   h c$, where the new physics 
effect is treated as an effective interaction between the top quark, the charm quark and the Higgs boson.

This paper is organised as follows: in section~\ref{sec:tbwz} we consider theoretical aspects of the $t \rightarrow bWZ$ decay 
and construct a simple phenomenological analysis to assess its observability at the FCC-hh within the SM. To complement this analysis, we study the allowed size of an enhancement to the decay rate due to the presence of a charged heavy scalar boson contribution, given current experimental constraints. Subsequently, in section~\ref{sec:thc}, we analyse the decay $t \rightarrow h c$, taking into account the effective coupling $h$-$t$-$c$ and extract a bound by looking at a clean final state at the FCC-hh, with $h\rightarrow \gamma\gamma$. Finally, we present our conclusions in section~\ref{sec:conclusions}. 

\section{\boldmath Top quark decays to $bWZ$}\label{sec:tbwz}
\subsection{Theoretical considerations}
\subsubsection{Defining the final state}
%We reproduce some of the results of previous publications 

The `cleanest' channels in which the process $t \rightarrow bWZ$ contributes are those containing multiple leptons. Here we will focus on the cases $t \rightarrow b e^+ \nu_e \mu^+\mu^-$ and $t \rightarrow b jj \mu^+\mu^-$, which receive contributions from other intermediate states in addition to $t \rightarrow bWZ$. We examine the observability of these processes by looking at the particle content in the final states only and we do not attempt to separate the $t \rightarrow bWZ$ contribution.\footnote{This separation is not possible due to interference of the $t \rightarrow bWZ$ contribution with other diagrams.}  

In Ref.~\cite{Altarelli:2000nt}, the ratio $R = \mathrm{BR}(t \rightarrow b \mu \nu_\mu \nu \bar{\nu})/[\mathrm{BR}(W\rightarrow  \mu \nu_\mu) \times \mathrm{BR}(Z \rightarrow \nu \bar{\nu})]$ was considered as a definition of the process $t \rightarrow bWZ$. Taking into account that the top quarks are on-shell, we can use an equivalent definition: 
\begin{equation}
R' = \frac{\mathrm{BR}\Bigl(pp \rightarrow t\bar{t} \rightarrow (b \mu^+ \nu_\mu \nu\bar{\nu}) \bar{t}\Bigl)}{\Bigl[\mathrm{BR}\Bigl(pp \rightarrow t\bar{t} \rightarrow (b \mu^+ \nu_\mu) \bar{t} \Bigl) \times \mathrm{BR}(Z \rightarrow \nu \bar{\nu})\Bigl]}.
\end{equation}
We have calculated $R'$ by using the \texttt{MG5\_aMC@NLO} Monte Carlo event generator~\cite{Alwall:2011uj, Alwall:2014hca}. Within the given errors, the results for $R'$ are in good agreement with those for the ratio $R$ appearing in the third column of Table~2 of Ref.~\cite{Altarelli:2000nt}. To make a direct comparison, we show the results in Table~\ref{tb:BRBWZ}, where we have used the values of the masses and constants given in Table 1 of Ref.~\cite{Altarelli:2000nt}.

\begin{table}[!htb]
  \begin{center}
    \begin{tabular}{ccc}
       $m_t$~[GeV]\ & $R'$ & $R$ (Ref.~\cite{Altarelli:2000nt}) \\ \hline
      170 & $1.55 \times 10^{-6}$ & $1.53(4)\times 10^{-6}$ \\
      171 & $1.62 \times 10^{-6}$ &  -\\
      172 & $1.71 \times 10^{-6}$ &  -\\
      173 & $1.79 \times 10^{-6}$ &  -\\
      174 & $1.89 \times 10^{-6}$ &  -\\
      175 & $2.00 \times 10^{-6}$ & $1.96(5)\times 10^{-6}$\\\hline
    \end{tabular}
  \end{center}
  \caption{The ratio of branching ratios defines the $t \rightarrow bWZ$ as described in the main text. The results of Ref.~\cite{Altarelli:2000nt} are only provided in 5~GeV intervals.}
\label{tb:BRBWZ}
\end{table}
\subsubsection{Next-to-leading order corrections}

\begin{figure}[!htb]
  \centering
    \includegraphics[width=1.0\linewidth]{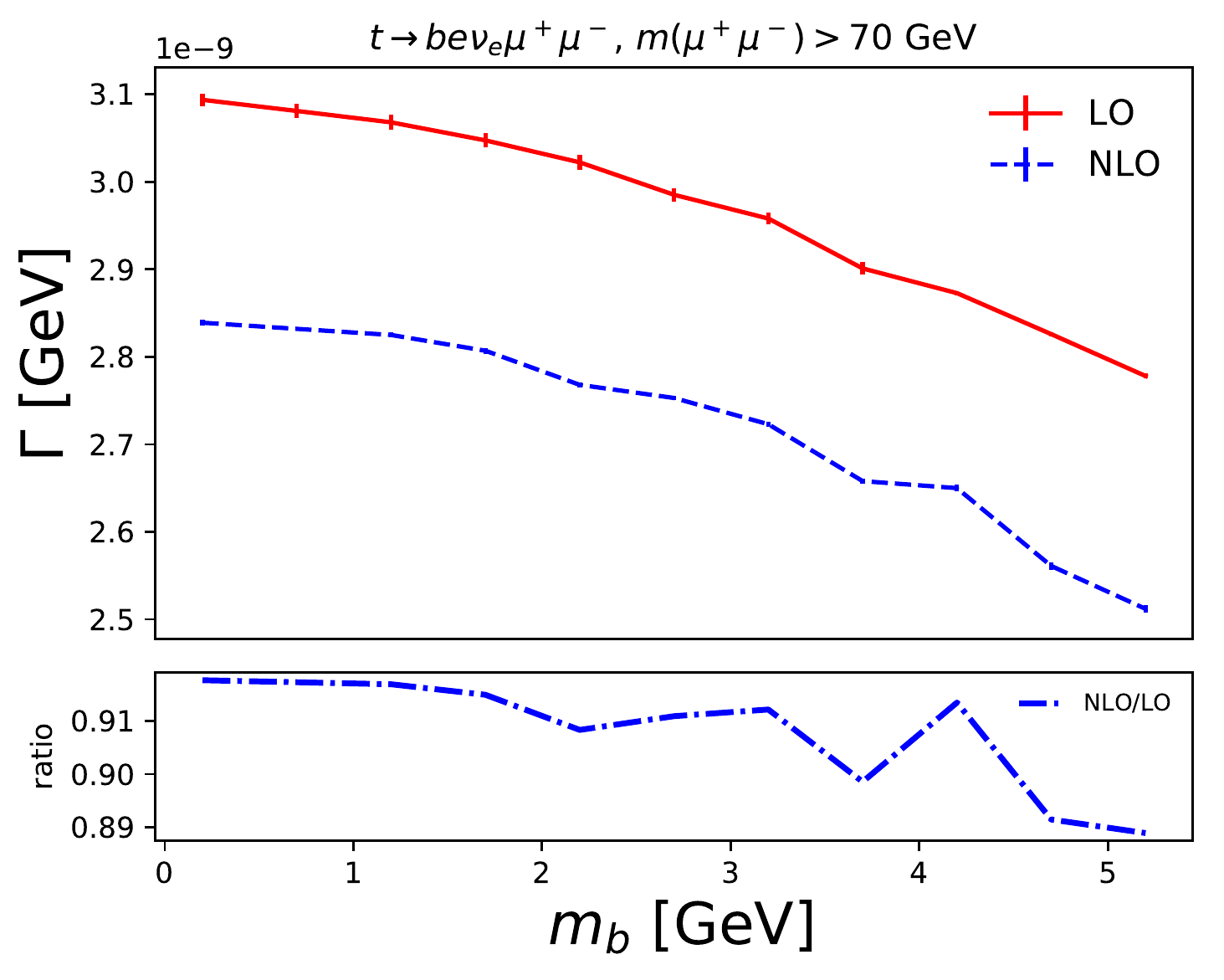}
  \caption{The decay width of the process $t \rightarrow b e^+ \nu_e \mu^+ \mu^-$ for $m(\mu^+\mu^-) > 70$~GeV at leading order and next-to-leading order in QCD as a function of the bottom quark mass. The lower inset shows the ratio of NLO to LO.}
  \label{fig:nlogamma}
\end{figure} 

Since the decay process occurs close to the top mass threshold, it is interesting to investigate the impact of next-to-leading order (NLO) 
QCD corrections. To consider a scenario which might be realistic in a phenomenological study, we examine corrections to the 
decay process $t \rightarrow b e^+ \nu_e \mu^+ \mu^-$, with the invariant mass of $\mu^+\mu^-$ pair taken to lie above $70$~GeV to 
remove the photon contribution, $\gamma \rightarrow \mu^+\mu^-$. We use \texttt{MG5\_aMC@NLO} to calculate the SM decay width, 
$\Gamma$, setting $m_t = 173$~GeV and varying the bottom mass between 0.2 and 5.2 GeV. In Figure~\ref{fig:nlogamma} we show the variation 
of the decay width as a function of the input bottom mass. In addition, we also present the ratio between the NLO 
and the leading order (LO) corrections in the lower inset. It is evident that the NLO corrections reduce the branching ratio by  about 10\%. 
The impact of increasing the $b$-quark mass is larger at NLO than at LO, which is due to the phase space becoming even more restricted close to the top mass threshold. Given the null results of the phenomenological analyses (see below), we do not consider higher-order corrections in more detail. 

\subsection{\boldmath Phenomenological analysis for $t\rightarrow bWZ$}
We construct a simple phenomenological analysis to determine whether a FCC-hh at 100~TeV will be sensitive to the SM top radiative 
decay $t \rightarrow bWZ$. We focus on the final states that arise through top pair production that contain one ``signal'' top decaying 
into 3 leptons and the other top decaying fully hadronically, i.e.
\begin{equation}
pp \rightarrow t\bar{t} \rightarrow (b \ell^{'+} \nu_\ell \ell^{+} \ell^{-})(\bar{b} jj)\;,
\end{equation}
and its charge-conjugate, where $j$ is any light-flavour jet. In addition
we also consider the case where one of the top quarks decays to oppositely-charged leptons and the other to only hadronic products:
\begin{equation}
pp \rightarrow t\bar{t} \rightarrow (b jj\mu^{+} \mu^{-})(\bar{b}  jj)\;,
\end{equation}
In our analysis, the jet reconstruction is done using the \texttt{fastjet} package~\cite{Cacciari:2005hq, Cacciari:2011ma}. We simulate the signal and backgrounds using \texttt{MG5\_aMC@NLO} at parton level and the general-purpose Monte Carlo event generator \texttt{HERWIG 7}~\cite{Gieseke:2011na, Arnold:2012fq, Bellm:2013lba, Bellm:2015jjp, Bellm:2017bvx} for the parton shower and non-perturbative effects, such as hadronization and multiple-parton interactions. 
 
\subsubsection{The three-lepton final state}
As a starting point for the analysis of the three-lepton final state:\footnote{Charge-conjugate processes are taken into account from this point on by multiplying by the appropriate symmetry factors.}
\begin{equation}
pp \rightarrow t\bar{t} \rightarrow (b \ell^{'+} \nu_\ell \ell^{+} \ell^{-})(\bar{b} jj)\;,
\end{equation}
we ask for two opposite-sign same-flavour leptons and one additional lepton, with $p_T > 20$~GeV within a pseudo-rapidity of $|\eta| < 6$. Furthermore, we ask that the opposite-sign same-flavour leptons reconstruct the $Z$ mass, $m_{\ell^+ \ell^-} \in [89, 93]$~GeV. As before, jets are reconstructed with the anti-$k_\perp$ algorithm with $R=0.3$ and we demand that they satisfy $p_T > 20$~GeV within $|\eta| < 6$. Two of these jets are $b$-jet candidates. Here we assume that $b$-jets can be tagged with 100\% efficiency, so as to give an upper estimate of the sensitivity to the processes under consideration. We then find the best combination of two \textit{or} three jets that reconstruct the top quark mass by looking at all the combinations of jets. We call this combination the ``reconstructed hadronic top'' and require $m_{t,\mathrm{reco}} \in [153, 193]$~GeV. We also ask that the distance $\Delta R$ between\footnote{Defined as usual $\Delta R = \sqrt{\Delta \eta^2 + \Delta \phi ^2}$, where $\eta$ is the pseudo-rapidity and $\phi$ is the azimuthal angle.} the sub-leading $b$-jet and the reconstructed $Z$ boson is $\Delta R < 1.0$. 

Using the assumption that the missing transverse momentum is primarily due to the undetected neutrino from the ``signal'' top quark decay and using the mass-shell condition $m_t^2 = (p_\nu + p_{\ell^{'+}} + p_{\ell^{+}} +  p_{\ell^{-}} + p_b)^2$ we obtain a quadratic equation, and hence two solutions, for the $z$-component of the neutrino momentum. We use these solutions to reconstruct two corresponding values of the total invariant mass using momenta of the reconstructed top quarks. We ask require \textit{both} of these values lie within $[ 350, 700]$~GeV. 

To assess the detection prospects of this process we consider the background arising from $pp \rightarrow t\bar{t} Z$, where all three particles are taken to be on-shell in this case. Using the aforementioned basic cuts, and the LO cross sections for the signal, $\sigma_{\mathrm{signal}} \simeq 3.00 \times 10^{-5}$~pb, and for the background contribution to the final state considered here (i.e. including branching ratios) $\sigma_{t\bar{t} Z} \simeq 0.10$~pb, we find an estimate of $\mathcal{O}(5)$ events for the signal and $\mathcal{O}(5000)$ events for the $t\bar{t} Z$ at an integrated luminosity of 10~ab$^{-1}$, a ballpark estimate of the FCC-hh end-of-lifetime data sample. This implies that this channel will be impossible to observe during the FCC-hh lifetime, and we do not consider it here any further. 

\subsubsection{The two-lepton final state}

To further investigate top quark radiative decays, let us now consider the signal process:
\begin{equation}
pp \rightarrow t\bar{t} \rightarrow (b jj\mu^{+} \mu^{-})(\bar{b}  jj)\;.
\end{equation}
As before, we consider only the main background channel: $pp \rightarrow t\bar{t}  Z \rightarrow (b j j)(\bar{b}  jj) (\mu^+ \mu^-)$, where the pair of muons arises from the decay of an on-shell $Z$ boson. Our selection procedure focuses on the reconstruction of the top quark based on the combination of final state products $(b jj\mu^{+} \mu^{-})$. To reconstruct the $Z$ boson we require two oppositely-charged muons satisfying $|\eta|<6$ such that the combined invariant mass is within the interval $[80,100]~\hbox{GeV}$. We cluster final state particles, excluding the muons, with $p_T > 5~\hbox{GeV}$  into anti-$k_{\perp}$ jets of  $R=0.3$. We require exactly two $b$-tagged jets.  We then recluster the constituents of the $R=0.3$ subjets into $R=1.2$ ``fat jets'' using the anti-$k_{\perp}$ algorithm, excluding the previously $b$-tagged subjets. For each fat jet, we apply a mass drop algorithm as in Ref.~\cite{Butterworth:2008iy} with parameters $y_{cut}=1.5$ and $\mu=0.25$.\footnote{We employ this method to ``groom'' the jets, removing soft radiation.} We determine the invariant mass $M'$  of the system composed by the pair of muons that reconstruct the $Z$ boson, the lowest $p_{T}$ $b$-tagged jet and the resulting subjet after the mass-drop  application. The event is
selected only if there is a fat jet such that  $140~\hbox{GeV} < M'<180~\hbox{GeV}$ after the mass drop conditions are applied. Based on this selection we obtain the cross sections $\sigma_{signal}=1.4\times 10^{-4}~\hbox{pb}$  and $\sigma_{background}=0.5~\hbox{pb}$, for signal and background respectively, corresponding to a significance $S/\sqrt{B}=0.6$ at an integrated luminosity of 10~ab$^{-1}$. Thus, it seems unlikely that this process would be detected in this channel.

\subsection{Heavy charged Higgs bosons}

\begin{figure}[!htb]
  \centering
    \includegraphics[width=0.48\linewidth]{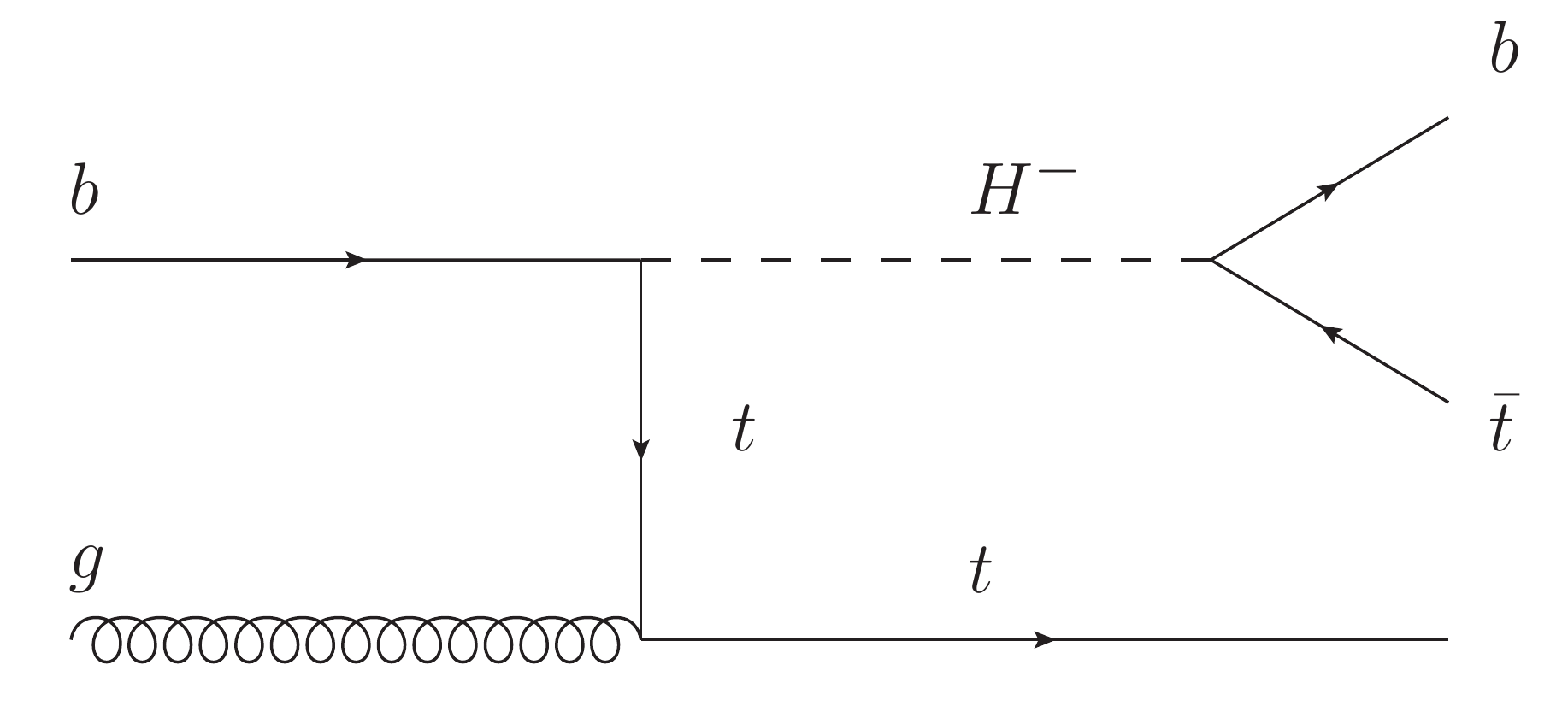}
    \includegraphics[width=0.48\linewidth]{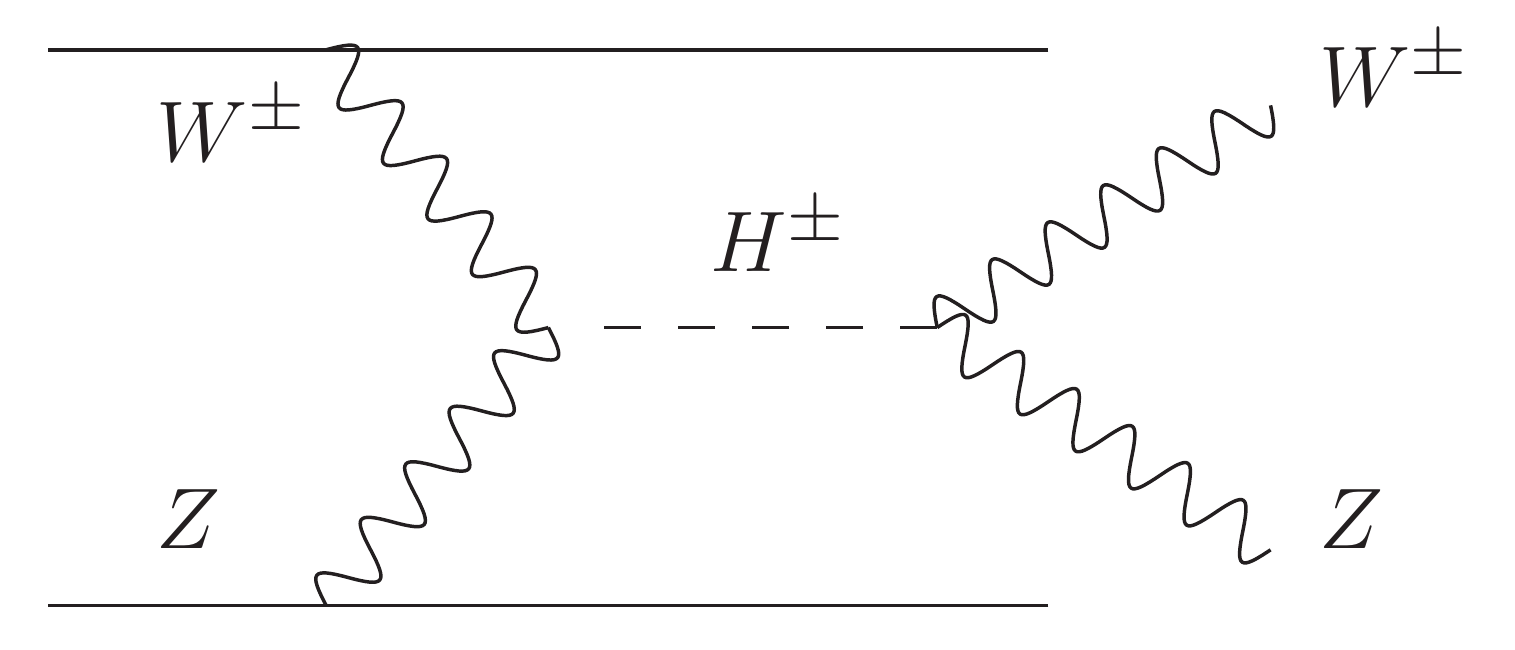}
  \caption{The heavy charged Higgs process $gb \rightarrow  H^- t \rightarrow (\bar{t}b)t$ (left, charge conjugate implied) and VBF to $H^\pm \rightarrow W^\pm Z$ (right) related to the ATLAS experimental analyses at 8~TeV~\cite{Aad:2015typ, Aad:2015nfa}, used here to extract the limits on the $H^\pm-t-b$ and $H^\pm - W^\pm - Z$ couplings, respectively.}
  \label{fig:heavyhdiagrams}
\end{figure} 

The analyses of the previous sections indicate that the SM radiative decay of the top quark seems challenging to detect even at the 100 TeV FCC-hh. Nevertheless, one might ask whether it would be possible to observe an enhanced rate due to beyond-the-SM contributions. One possibility arises with the addition of a heavy charged Higgs boson, $H^\pm$ that couples to $W^\pm Z$ as well as top and bottom quarks. Such couplings have been probed in LHC experimental analyses, for example the ATLAS analyses at 8 TeV that appear in Refs.~\cite{Aad:2015typ, Aad:2015nfa}. In Ref.~\cite{Aad:2015typ}, the process $gb \rightarrow  H^- t \rightarrow (\bar{t}b)t$ (and charge conjugate) was searched for by the ATLAS experiment, using data corresponding to an integrated luminosity of 20.3 fb$^{-1}$. The diagram contributing to this channel, shown on the left in Fig.~\ref{fig:heavyhdiagrams}, has a rate that is directly proportional to the quartic power of the $H^\pm-t-b$ coupling. Similarly, in Ref.~\cite{Aad:2015nfa}, the final state $H^\pm \rightarrow W^\pm Z$ was searched for using an equivalent dataset in vector boson fusion. The latter process, shown on the right in Fig.~\ref{fig:heavyhdiagrams}, has a rate proportional to the quartic power of the $H^\pm -  W^\pm - Z$ coupling. 

\begin{figure}[!ht]
  \centering
    \includegraphics[width=1.0\linewidth]{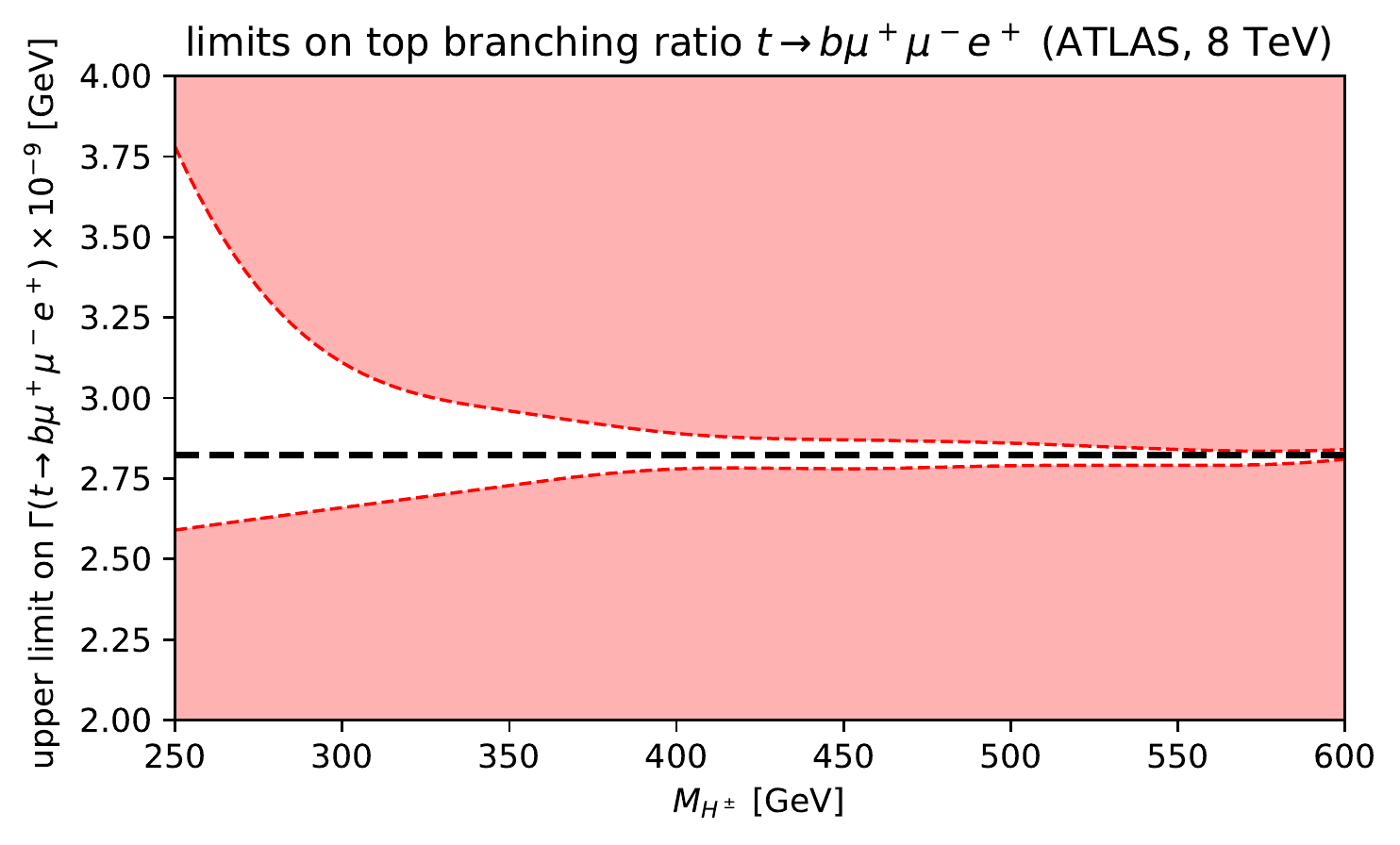}
  \caption{The 95\% C.L. limits on the decay width of the process $t \rightarrow b e^+ \nu_e \mu^+ \mu^-$ for $m(\mu^+\mu^-) > 70$~GeV at leading order through a hypothetical heavy charged scalar that couples only to $W^\pm Z$ and top and bottom quarks. The constraints on the couplings were obtained from the ATLAS experimental analyses at 8~TeV~\cite{Aad:2015typ, Aad:2015nfa}.}
  \label{fig:twzbheavyh}
\end{figure} 

We simulate these two processes at LO using \texttt{MG5\_aMC@NLO} and assume that the new scalar \textit{only} possesses these two interactions. Hence, using the results obtained in the aforementioned articles~\cite{Aad:2015typ, Aad:2015nfa}, we derive constraints for the maximum and minimum allowed values of the decay width at LO. These are shown in Fig.~\ref{fig:twzbheavyh}. Evidently the enhancement factor is moderate over the range of scalar boson masses considered, with a maximal value of $\mathcal{O}(2)$ for a heavy charged Higgs boson mass of $\sim 200$~GeV. Hence we can conclude that the addition of a heavy charged scalar cannot render this process observable at a 100~TeV collider. Note that due to the interference of the SM diagrams with the charged scalar diagrams, which can be negative, the decay width can possess values lower than those of the SM. 

\section{\boldmath Top quark decays to Higgs boson-charm quark}\label{sec:thc}

\subsection{\boldmath The $h$-$t$-$c$ coupling}

We now turn to the investigation of flavour-changing neutral decays of the top quark. We consider gauge-invariant and renormalizable Yukawa interactions of the form 

\begin{equation}
\mathcal{L}_{thc} = \lambda_{ct}^h \bar{Q}_c H q_t + \mathrm{h.c.}\;,
\end{equation}\label{eq:htc}
where $Q_i$ is a left-handed doublet, $q_j$ is a right-handed singlet and $H$ is the SM Higgs doublet. We constrain our analysis to real and symmetric couplings: $\lambda_{ct}^h = \lambda_{tc}^h = (\lambda_{ct}^h)^\dagger = (\lambda_{tc}^h)^\dagger$. 

Here we focus on the resulting top quark decay $t \rightarrow h c$, induced by couplings of the above kind. We also note that these couplings 
can lead to other interesting final states~\cite{Atwood:2013ica}.  Various studies have already examined this process at the LHC~\cite{AguilarSaavedra:2000aj, Craig:2012vj, Atwood:2013ica}, with the current best experimental constraint on BR($t\rightarrow hc$) at the LHC being $0.22\%$ at 95\% C.L., coming from ATLAS 13 TeV data (36.1 fb$^{-1}$) in the di-photon channel~\cite{Aaboud:2017mfd}. The corresponding best constraint at CMS is currently $0.47\%$ through $h\rightarrow b \bar{b}$ decays~\cite{Sirunyan:2017uae}. Naive extrapolation of the 13 TeV ATLAS result~\cite{Aaboud:2017mfd} to the high-luminosity LHC demonstrates that an integrated luminosity of 3000~fb$^{-1}$ implies an ultimate constraint of BR($t\rightarrow hc$)~$\lesssim 0.019\%$ through the $h\rightarrow \gamma\gamma$ channel alone.\footnote{This is obtained by extrapolating the number of events for the signal and backgrounds from 36.1~fb$^{-1}$ to 3000~fb$^{-1}$, assuming that the experimental details and analysis remain unchanged.} It is important to note here that the LHC analyses do not consider the tagging of charm jets in the derivation of these constraints. This implies that these limits are associated with the crucial assumption that the $t \rightarrow h u$ decay will be either absent or sub-dominant with respect to $t \rightarrow h c$. Alternatively, one can use these analyses to impose constraints on $t \rightarrow h u$, assuming $t\rightarrow hc$ is absent or sub-dominant.

In the present study we will analyse the prospect of constraining $\mathrm{BR}(t \rightarrow h c)$ and $\lambda_{ct}^h$ through top quark pair production at the FCC-hh. To the best of our knowledge this represents the first estimate for a constraint on this coupling at the FCC-hh. Among all the decay channels, the one expected to provide the strongest contribution in the combination for the constraint is the one involving the transition $h \rightarrow \gamma \gamma$, and therefore we will focus on it in the present study. In our analysis we consider both the scenario with and that without charm-jet tagging, with values for the tagging efficiencies motivated by current LHC considerations~\cite{ATL-PHYS-PUB-2015-001}. 

\subsection{\boldmath Phenomenological analysis for $t\rightarrow hc$}

\begin{table}[!htb]
  \begin{center}
    \begin{tabular}{lcc}
       Process & $\sigma_\mathrm{gen}^{\mathrm{had.}}$ [pb] & $\sigma_\mathrm{gen}^{\mathrm{s.l.}}$ [pb] \\ \hline
       $pp \rightarrow t\bar{t} \rightarrow (hc) \bar{t} +$ h.c. 	& 0.332 & 0.122 \\
      $pp \rightarrow t \bar{t} h$ & 0.044 & 0.030 \\
      $pp \rightarrow h j j W^\pm$ & 0.022 & 0.070 \\
      $pp \rightarrow t \bar{t} \gamma \gamma$ & 0.042 & 0.028 \\ 
      $pp \rightarrow \gamma \gamma j j W^\pm$ & 1.294 & 0.424 \\\hline
    \end{tabular}
  \end{center}
  \caption{The starting signal and background cross sections considered in the analyses of the top quark pair production search for the $t\rightarrow hc$ decay. For simplicity, we have rescaled the leading-order cross sections for all processes by a $k$-factor of 2. This approximation does not have a significant impact on our conclusions. The second and third columns show the generation-level cross sections for the hadronic and semi-leptonic cases, respectively, see main text for further details. The signal cross sections are shown for $\lambda^h_{ct} = 0.1$, which we take here as a ``working value''.}
\label{tb:thcxsecs}
\end{table}

We have implemented the interaction described by Eq.~(\ref{eq:htc}) in
a UFO~\cite{Degrande:2011ua} model which we interface to
\texttt{MG5\_aMC@NLO} to generate signal $pp \rightarrow t\bar{t}
\rightarrow (hc) \bar{t}$ (and the charge-conjugate process)
events. We also generate parton-level events for the backgrounds using
\texttt{MG5\_aMC@NLO} and perform shower and hadronization using
\texttt{HERWIG 7} as before. The background processes considered
include those that include a Higgs boson in association with other
particles: $pp \rightarrow t \bar{t} h$, $pp \rightarrow h j j W^\pm$
and those that contain non-resonant di-photon production: $pp
\rightarrow t \bar{t} \gamma \gamma$, $pp \rightarrow \gamma \gamma j
j W^\pm$, where the $W$ bosons were decayed to electrons or
muons. Generation-level cuts were applied on the non-resonant photon
samples: the photon transverse momentum was required to lie in
$p_{T,\gamma} > 10$~GeV, the distance between either a jet and a
photon or between two photons $\Delta R (\gamma, \mathrm{~j~or}~\gamma) > 0.1$
and the invariant mass of the two photons to satisfy $M_{\gamma\gamma} \in [110, 140]$~GeV. In all background samples we asked for the generation-level cuts on the jets and final-state leptons of $p_T > 20$~GeV. The jets have been merged to the \texttt{HERWIG 7} parton shower at tree level using the MLM method via the FxFx add-on module~\cite{Frederix:2015eii}.\footnote{Further details on the usage of this module for tree-level merging will available in a future release of the \texttt{HERWIG 7} manual.} 

Our analysis is divided into two cases, either with the ``non-signal'' top decaying hadronically, ($t \rightarrow b jj$) or semi-leptonically ($t \rightarrow b \ell \nu$). The starting cross sections for both the hadronic and semi-leptonic final states, with the generation-level cuts described above, are given in the second column of Table~\ref{tb:thcxsecs}. For the signal cross section we use a working value of $\lambda^h_{ct} = 0.1$. To take into account higher-order effects and for the sake of simplicity, we have rescaled the leading-order cross sections for all processes by a $k$-factor of 2. This approximation does not have a significant impact on our derived constraints and can be fully addressed in a future analysis.

We identify photons and leptons by requiring $p_T > 25$~GeV, within $|\eta| < 2.5$ in both cases. We assume flat identification efficiencies of $b$-jets of 70\% and of $c$-jets of 20\% and ask for them to have $p_T > 25$~GeV and lie within $|\eta| < 2.5$~GeV.\footnote{In general charm-jet tagging is currently less developed than $b$-jet tagging, see e.g.~\cite{ATL-PHYS-PUB-2015-001} for single-prong charm-jet tagging algorithm and~\cite{Lenz:2017lqo} for a double-charm-jet tagger used in the context of Higgs searches.} We consider mis-tagging rates for light jets to $b$-jets of 1\%, and to $c$-jets of 0.5\%. The rate of mis-identification of $b$-jets to $c$-jets was taken to be 12.5\% and the rate for the converse was taken to be 10\%~\cite{ATL-PHYS-PUB-2015-001}. We do not consider mis-tagging of light jets to photons in our analysis. We do not apply any detector effects such as momentum smearing and we assume that jets, leptons and photons are detected with 100\% efficiency within the considered coverage.\footnote{As discussed in, e.g.,~\cite{Papaefstathiou:2015paa}, better forward detector coverage for $b$-jet or photon identification, up to $|\eta|\sim 3-3.5$ may increase signal efficiency at a future 100 TeV collider. In the present analysis we chose to be conservative, allowing identified objects only within $|\eta| < 2.5$.}

In all cases, we ask for exactly one identified $b$-jet and at least two photons. For the semi-leptonic top case we ask for at least one lepton. We reconstruct the signal top quark from the identified $b$-jets and di-photon system and ask for the mass to lie within $m_{\gamma \gamma c} \in [160, 190]$~GeV. Furthermore, we ask for the di-photon invariant mass to reconstruct the Higgs boson mass within 2 GeV: $m_{\gamma\gamma} \in [123, 127]$~GeV, the distance between the photons to lie within $\Delta R (\gamma,\gamma) \in [1.8, 5.0]$ and the distance between the di-photon system and the $c$-jet to lie within $\Delta R(\gamma\gamma, c) < 1.8$. In the case of no charm-jet tagging we simply consider all non-$b$-jets candidates as light. In practice, this amounts to not having a specific tagging weight when considering ``true'' charm jets. In the semi-leptonic top case we assume that the missing transverse energy is entirely due to the missing neutrino and reconstruct its $z$-component by solving the quadratic equation $m_W^2 = (p_\ell + p_\nu)^2$, where $p_\ell$ and $p_\nu$ represent the 4-momenta of the hardest lepton and the missing neutrino respectively. Here, we take $m_W = 80.4$~GeV. We then ask for one of the two solutions to reconstruct the top mass when combined with the $b$-jet within the range $[150, 200]$~GeV. In the hadronic top case we consider the invariant mass of combinations of the $b$-jet with one \textit{or} two light jets and find the one closest to the top mass, taking $m_\mathrm{top} = 173$~GeV. We then ask for this to lie in the same range: $[150, 200]$~GeV.

\begin{table}[!htb]
\centering
\setlength\extrarowheight{2pt}
\begin{tabularx}{0.5\textwidth}{|*{3}{>{\centering\arraybackslash}X|}}
  \hline
  \multicolumn{3}{c}{exactly one $b$-jet, $p_T > 25$~GeV, $|\eta| < 2.5$,}\\
  \multicolumn{3}{c}{$P_{b\rightarrow b} = 0.7$, $P_{c\rightarrow b} = 0.1$, $P_{l\rightarrow b} = 0.01$,}\\
  \multicolumn{3}{c}{$\geq 2$ photons, $p_T > 25$~GeV, $|\eta| < 2.5$,}\\\hline
  \multicolumn{3}{@{}c@{}}{\begin{tabularx}{\dimexpr 0.5\textwidth-2\arrayrulewidth}[t] {>{\centering\arraybackslash}X|>{\centering\arraybackslash}X}
    \textbf{hadronic:} & \textbf{semi-leptonic:}\\
    $\geq 1$ light jets, & $\geq 1$ leptons, $p_T > 25$~GeV, $|\eta| < 2.5$, \\
    top: combine $b$-jet + 1, 2 light jets. & solve for $p^z_\nu$ using mass constraint. \\
    \hline
    \textbf{with $c$-tagging:} & \textbf{no $c$-tagging:}\\
     $P_{c\rightarrow c} = 0.2$, $P_{l\rightarrow c} = 0.005$, $P_{b\rightarrow c} = 0.125$.& \vspace{0.04cm}no charm jets.\\
  \end{tabularx}} \\\hline
  \multicolumn{3}{c}{$m_\mathrm{top,~reco} \in [150, 200]$~GeV.}\\
  \multicolumn{3}{c}{$m_{\gamma\gamma c} \in [160, 190]$~GeV,}\\
  \hline
\end{tabularx}
\caption{A summary of the selection criteria of the analysis for each of the channels considered for the $pp \rightarrow t\bar{t} \rightarrow (hc) \bar{t}$ process. The final invariant mass cut, on $m_{\gamma\gamma c}$ allows identification of the signal top quark.}
\label{tb:thcanalysisdescription}
\end{table}

\begin{table}[!htb]
  \begin{center}
    \begin{tabular}{lcc}
      \hline
     \multicolumn{3}{c}{ $\mathcal{L} = 10~\mathrm{ab}^{-1}$  }  \\\hline
       Process & $N_\mathrm{c-tag}^{\mathrm{had.}}$ & $N_\mathrm{c-tag}^{\mathrm{s.l.}}$ \\ \hline
      $pp \rightarrow t\bar{t} \rightarrow (hc) \bar{t} +$ h.c. & 22952 & 10260 \\
      $pp \rightarrow t \bar{t} h$ & 1816 & 689 \\
      $pp \rightarrow h j j W^\pm$ & 7 & 1 \\
      $pp \rightarrow \gamma \gamma j j W^\pm$ &  211  & 2 \\ 
      $pp \rightarrow t \bar{t} \gamma \gamma$ & 107 & 39 \\\hline
    \end{tabular}
  \end{center}
  \caption{The expected signal and background events at an integrated luminosity of $\mathcal{L} = 10~\mathrm{ab}^{-1}$ after applying the
    analyses in the search for the $t \rightarrow hc$ decay. The
    resulting event yields are shown for the case where charm-jet tagging is considered for the hadronic and semi-leptonic cases, see main text for further details. As before, the signal cross sections are shown for the working value $\lambda^h_{ct} = 0.1$.}
\label{tb:thcxsecs_analysis_charm}
\end{table}

\begin{table}[!htb]
  \begin{center}
    \begin{tabular}{lcc}      
      \hline
     \multicolumn{3}{c}{ $\mathcal{L} = 10~\mathrm{ab}^{-1}$  }  \\\hline
       Process & $N_\mathrm{no~c-tag}^{\mathrm{had.}}$  & $N_\mathrm{no~c-tag}^{\mathrm{s.l.}}$ \\ \hline
      $pp \rightarrow t\bar{t} \rightarrow (hc) \bar{t} +$ h.c. & 191871 & 61124 \\
      $pp \rightarrow t \bar{t} h$ &  26533  & 6962\\
      $pp \rightarrow h j j W^\pm$ & 66  & 19 \\
      $pp \rightarrow \gamma \gamma j j W^\pm$ & 7130 & 164\\ 
      $pp \rightarrow t \bar{t} \gamma \gamma$ & 1598 & 478 \\\hline
    \end{tabular}
  \end{center}
  \caption{As for Table~\ref{tb:thcxsecs_analysis_charm}, but without
    charm-jet tagging.}
\label{tb:thcxsecs_analysis_nocharm}
\end{table}

We summarise the main features of the analysis in
Table~\ref{tb:thcanalysisdescription}. The resulting event yields after applying the analyses are shown in
Tables~\ref{tb:thcxsecs_analysis_charm} and~\ref{tb:thcxsecs_analysis_nocharm} for the cases with and without
charm tagging, respectively, at an integrated luminosity of $\mathcal{L} = 10~\mathrm{ab}^{-1}$.

\subsection{\boldmath Constraints for $t\rightarrow hc$}

To take into account the effect of the presence of systematic uncertainties, we assume that they only affect the total number of background events, $B$, by inducing a systematic uncertainty $\Delta B = \alpha B$, with $\alpha \geq 0$ parameterising the effect. We add this in quadrature to the statistical uncertainty on the expected number of events. We therefore show results for values of  $\alpha$ corresponding to no systematics ($\alpha=0$) to demonstrate the ultimate precision at the future collider, low systematic uncertainty ($\alpha = 0.05$), and high systematic uncertainty ($\alpha = 0.2$). For the ATLAS analysis of~\cite{Aaboud:2017mfd} we have deduced that the current systematic uncertainty would correspond to $\alpha \simeq 0.063$ and we derive results for an extrapolation to the high-luminosity LHC data set (3000~fb$^{-1}$) either using this value or setting $\alpha = 0$ as the best-case scenario. 

\begin{figure}[!htb]
  \centering
    \includegraphics[width=1.0\linewidth]{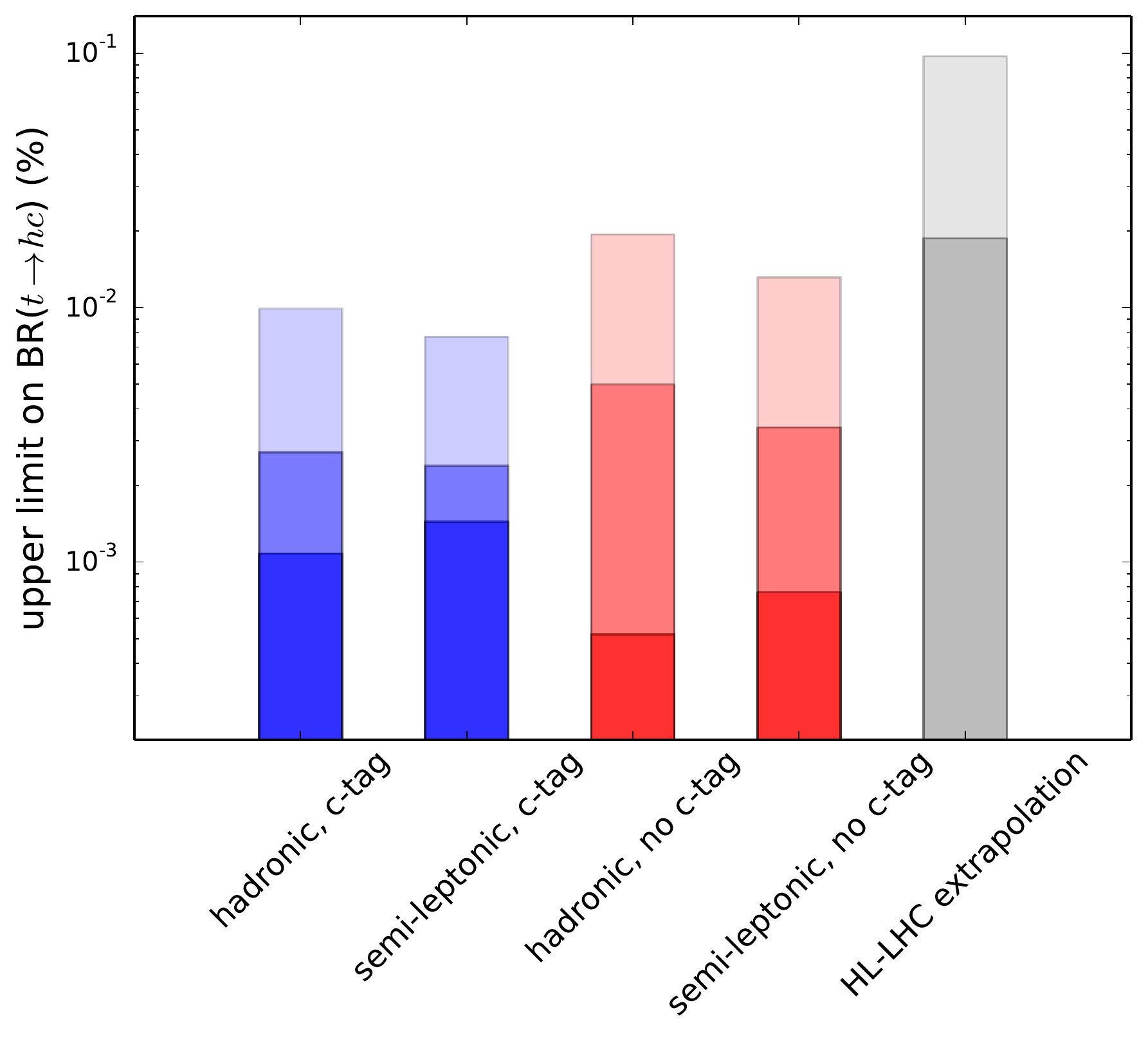}
  \caption{The upper limits on the branching ratio BR($t \rightarrow hc$) as a percentage of the total, as calculated by each of the phenomenological analyses of this article for a 100 TeV FCC-hh with an integrated luminosity of 10~ab$^{-1}$. The (blue and red) bars represent the $95\%$ C.L. limits taking into account systematic uncertainties $\alpha = (0, 0.05, 0.2)$ going from darker to lighter-shaded, respectively. The blue bars represent the cases with charm-jet tagging applied, whereas the red bars represent the cases without it. The darker grey bar represents a naive statistical extrapolation of the ATLAS constraints that appear in~\cite{Aaboud:2017mfd} to the full high-luminosity LHC data set (3000~fb$^{-1}$) in the absence of systematic uncertainties ($\alpha =0$) and the lighter-shaded grey area roughly takes into account the current estimate of the systematic uncertainties, corresponding to $\alpha \simeq 0.063$.}
  \label{fig:thcconstraints}
\end{figure} 

We show the resulting constraints on the percent branching ratio in Fig.~\ref{fig:thcconstraints}. The values for 95\% C.L. upper limits are also given in Table~\ref{tb:thcconstraints}. Both the analyses with and without charm-jet tagging are able to provide constraints on the branching ratio of $\mathcal{O}(10^{-3})\%$. A naive statistical combination of the hadronic and semi-leptonic channels yields 95\% C.L. upper limits on BR($t\rightarrow hc$) of $(8.5 \times 10^{-4})\%$ for the case with charm-jet tagging and $(4.4 \times 10^{-4}) \%$ without for the case of no systematics $(\alpha = 0$).\footnote{The naive statistical combination employed here adds the Gaussian significances linearly: $\sigma_\mathrm{total} = \sum_{i=1}^k \frac{\sigma_i}{\sqrt{k}}$. This provides a conservative estimate of the combined significance~\cite{stouffer}.} For the cases $\alpha = (0.05, 0.2)$ the combination yields: BR($t\rightarrow hc$) of $(1.8, 6.1) \times 10^{-3})\%$ for the case with charm-jet tagging and $(2.8, 11.1) \times 10^{-3}) \%$ without, respectively. We also show the extrapolation of the ATLAS constraints of~\cite{Aaboud:2017mfd}, which is at least an order of magnitude worse than the results of the present analysis for comparable systematics, corresponding to BR($t\rightarrow hc$) of $(1.9, 9.7) \times 10^{-2}\%$ for $\alpha = (0, 0.063)$ respectively.\footnote{These estimates are in agreement with the HL-LHC projections found in Ref.~\cite{ATLASHL}.} This would correspond to the case of no charm-jet tagging of the present analysis.\footnote{The increase in signal cross section from 13~TeV to 100~TeV is $\sim 40$, whereas for the $pp \rightarrow t\bar{t}h$ background this increase is $\sim 70$. Given that we are now considering 10~ab$^{-1}$ versus 3~ab$^{-1}$ at the HL-LHC, the naive increase in significance is $\sim ( 40/\sqrt{70}) \times \sqrt{10/3} \sim 9 $, which would imply an order of magnitude improvement on the branching ratio measurements from 13~TeV to 100~TeV, provided the kinematical structure scales in a similar way for signal and background.}
\begin{table}[!htb]
  \begin{center}
    \begin{tabular}{llll}
      \multicolumn{4}{c}{ with $c$-tagging: }   \\\hline
       \multicolumn{2}{l}{analysis:} & hadr.,  & semi-lept.\\ \hline
      $\lambda_{ct}^h$ $\times 10^{-3}$ & & (6.42, 10.15, 19.40)  & (7.40, 9.52, 17.08) \\
      BR in $10^{-3}$\%  & & (1.08, 2.70, 9.91) & (1.44, 2.39, 7.69) \\\hline
     \multicolumn{4}{c}{ no $c$-tagging: }  \\\hline
      $\lambda_{ct}^h$ $\times 10^{-3}$ &  & (4.43,  13.61, 27.15)  & (5.38, 11.32, 22.36)   \\
      BR in $10^{-3}$\%  & & (0.52, 4.99, 19.42)& (0.76, 3.38, 13.17)\\\hline

    \end{tabular}
  \end{center}
  \caption{The 95\% C.L. upper limits as calculated by each of the phenomenological analyses of this article for a 100 TeV FCC-hh with an integrated luminosity of 10~ab$^{-1}$ corresponding to the systematic uncertainty parameter values $\alpha = (0, 0.05, 0.2)$ as described in the main text. We give the limits on the branching ratios for the top quark as a percentage of the total as well as the associated values of the coupling, $\lambda_{ct}^h$. The results for BR($t \rightarrow hc$) are also given graphically in Fig~\ref{fig:thcconstraints}.}
\label{tb:thcconstraints}
\end{table}
\section{Conclusions}\label{sec:conclusions}
We have investigated the rare top quark decay processes $t \rightarrow bWZ$ and $t \rightarrow h c$ at a future circular hadron collider running at 100 TeV with 10~ab$^{-1}$ of integrated luminosity.  We have demonstrated that it will be extremely challenging to observe a final state in which the $t \rightarrow bWZ$ process contributes. This is true even in the case of the presence of new physics contributions allowed by current LHC constraints. On the other hand, the $t \rightarrow h c$ decay can be constrained to $\mathcal{O}(10^{-3})\%$, either with or without considering charm-jet tagging. This estimate is an order of magnitude more stringent than a high-luminosity LHC extrapolation and will allow us to constrain the off-diagonal top quark-charm quark Yukawa couplings to $\lambda_{ct}^h \sim \mathcal{O}(10^{-3})$.

The extremely rare decay modes we have investigated in the present article constitute two of the many interesting ones for top quarks. A future high-energy collider will be able to provide information on other processes, such as $t \rightarrow cWW$, $t \rightarrow q \gamma$, $t \rightarrow q Z$, $t \rightarrow c \gamma \gamma$ and $t \rightarrow c ZZ$. We leave investigations of such modes to future work.  

\section*{Acknowledgements}
We would like to thank Michael Spannowsky for useful discussions. AP acknowledges support by the ERC grant ERC-STG-2015-677323.

\providecommand{\href}[2]{#2}\begingroup\raggedright\endgroup

\end{document}